\documentclass{llncs}
\usepackage[utf8]{inputenc}

\title{Level-Planar Drawings with Few Slopes}
\author{Guido Br\"uckner \and Nadine Davina Krisam \and Tamara Mchedlidze}
\institute{Karlsruhe Institute of Technology\\
          \email{brueckner@kit.edu, nadine.krisam@student.kit.edu, mched@iti.uka.de}}

\usepackage{amsmath}
\usepackage{amsfonts}
\usepackage{graphicx}

\DeclareMathOperator{\high}{high}
\DeclareMathOperator{\low}{low}
\DeclareMathOperator{\Left}{left}
\DeclareMathOperator{\Right}{right}

\binoppenalty=\maxdimen
\relpenalty=\maxdimen

\newcommand{\ldr}{$\lambda$-\textsc{Drawability}~}
\newcommand{\refv}{\operatorname{ref}}

\begin{document}

%\pagenumbering{gobble}

\maketitle

\begin{abstract}
    We introduce and study level-planar straight-line drawings with a fixed number~$\lambda$ of slopes.
    For proper level graphs, we give an $O(n \log^2 n / \log \log n)$-time algorithm that either finds such a drawing or determines that no such drawing exists.
    Moreover, we consider the partial drawing extension problem, where we seek to extend an immutable drawing of a subgraph to a drawing of the whole graph, and the simultaneous drawing problem, which asks about the existence of drawings of two graphs whose restrictions to their shared subgraph coincide.
    We present~$O(n^{4/3} \log n)$-time and~$O(\lambda n^{10/3} \log n)$-time algorithms for these respective problems on proper level-planar graphs.

    We complement these positive results by showing that testing whether non-proper level graphs admit level-planar drawings with~$\lambda$ slopes is \textsf{NP}-hard even in restricted cases.
    \end{abstract}

%\newpage
\pagenumbering{arabic}

\section{Introduction}

Directed graphs explaining hierarchy naturally appear in multiple industrial and academic applications. Some examples include PERT diagrams, %facilitate the project management process -- the vertices represent tasks and directed edges indicate the order in which the tasks need to be implemented. 
UML component diagrams,
% facilitate the software engenering process -- the vertices represent the components of a software, the edges indicate that one component provides the services that another component requires.
%In humanities, 
text edition networks~\cite{faust},
% represent the editing process of a manuscript -- each vertex is a witness of a text and a directed edge indicates that one witness emerged from another. 
text variant graphs~\cite{JanickeGFTMS15}, %allow for analysis of textual variantion -- a vertex is a word, two words are connected if there is a version of the text, where one directly precedes another.  In biology,.... \todo{Philogenetic networks, neural networks}
philogenetic and
neural networks.
In these, and many other applications, it is essential to visualize the implied directed graph so that the viewer can perceive the hierarchical structure it contains. By far the most popular way to achieve this is to apply the \emph{Sugyiama framework} -- a generic network visualization algorithm that results in a drawing where each vertex lies on a horizontal line, called \emph{layer}, and each edge is directed from a lower layer to a higher layer~\cite{HealyN13}.

\begin{figure}[t]
 	\centering
 	\includegraphics[width=\columnwidth]{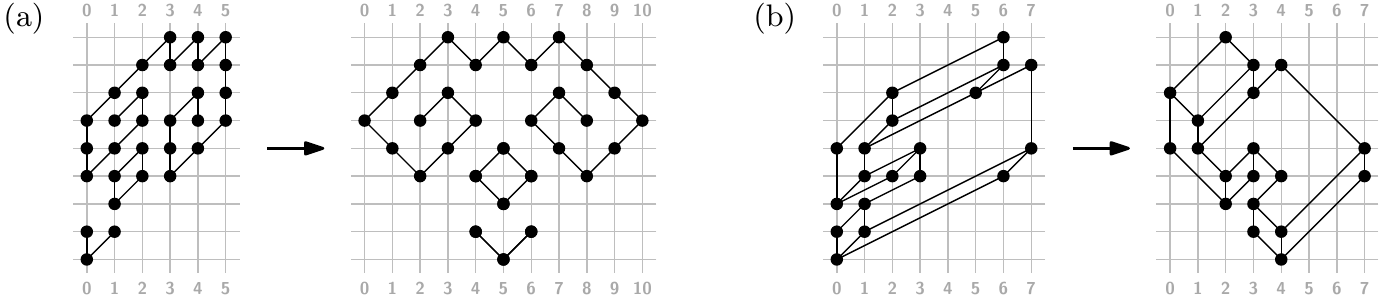}
 	\caption{%
 		Shearing drawings to change the slopes.
 		In~(a), the left drawing with slopes~$0$ and~$1$ is transformed into the right orthogonal drawing, i.e., one with slopes~$-1$ and~$1$.
 		In~(b), the left drawing with slopes~$0, 1$ and~$2$ is transformed into a drawing with slopes~$-1, 0$ and~$1$.
 	}
 	\label{fig:shearing}
 \end{figure}

The Sugyiama framework consists of several steps: elimination of directed cycles in the initial graph, assignement of vertices to layers, vertex ordering and coordinate assignement. During each of these steps several criteria are optimized, by leading to more readable visualizations, see e.g.~\cite{HealyN13}. % e.g. number of edges pointing down, the maximum width of the layer, number of crossing among the edges, edge straightness.  Depending on the exact problem formulation, many of these problems become NP-hard and multiple heuristic approaches for each of the steps have been proposed in literature. 
%We refer the reader to for the detailed description of the components of the framework. 
In this paper we concentrate on the last step of the framework, namely coordinate assignment.
Thus, the subject of our study are \emph{level graphs} defined as follows. Let~$G = (V, E)$ be a directed graph.
A \emph{$k$-level assignment} of~$G$ is a function~$\ell: V \to \{1, 2, \dots, k\}$ that assigns each vertex of~$G$ to one of~$k$ levels.
We refer to~$G$ together with~$\ell$ as to a ($k$-)\emph{level graph}.
The \emph{length} of an edge~$(u, v)$ is defined as~$\ell(v) - \ell(u)$.
We say that~$G$ is \emph{proper} if all edges have length one.
The level graph shown in Fig.~\ref{fig:shearing}~(a) is proper, whereas the one shown in~(b) is not.
For a non-proper level graph~$G$ there exists a \emph{proper subdivision} obtained by subdividing all edges with length greater than one which result in a proper graph.

A \emph{level drawing}~$\Gamma$ of a level graph~$G$ maps each vertex~$v \in V$ to a point on the horizontal line with~$y$-coordinate~$\ell(v)$ and a real~$x$-coordinate~$\Gamma(v)$, and each edge to a~$y$-monotone curve between its endpoints.
A level drawing is called \emph{level-planar} if no two edges intersect except in common endpoints.
It is \emph{straight-line} if the edges are straight lines.
A level drawing of a proper (subdivision of a) level graph~$G$ induces a total left-to-right order on the vertices of a level.
We say that two drawings are \emph{equivalent} if they induce the same order on every level.
An equivalence class of this equivalence relation is an \emph{embedding} of~$G$.
We refer to~$G$ together with an embedding to as \emph{embedded level graph}~$\mathcal G$.
%This definition of an embedding is a more restricted than that of a \emph{combinatorial embedding}, where two drawings are considered equivalent if the circular orders of the edges around each vertex are identical, see Figure~\ref{fig:embedding}.
The third step of Sugyiama framework, vertex ordering, results in an embedded level graph.
%In this paper we deal with embedded level-planar graphs and also assume that a level-planar drawing is straight-line and  \emph{preserves} the given embedding.

The general goal of the coordinate assignment step is to produce a final visualization while further improving its readability. The criteria of readability that have been considered in the literature for this step include
 straightness and verticality of the edges~\cite{HealyN13}.
 Here we study the problem of coordinate assignment step with bounded number of slopes.
 The \emph{slope} of an edge~$(u, v)$ in~$\Gamma$ is defined as~$(\Gamma(v) - \Gamma(u)) / (\ell(v) - \ell(u))$.
 For proper level graphs this simplifies to~$\Gamma(v) - \Gamma(u)$.  We restrict our study to drawings in which all slopes are non-negative; such drawings can be transformed into drawings with negative slopes by shearing; see Fig.~\ref{fig:shearing}.
 A level  drawing~$\Gamma$ is a~\emph{$\lambda$-slope drawing} if all slopes in~$\Gamma$ appear in the set~$\{0, 1, \dots, \lambda - 1\}$.
We study embedding-preserving straight-line level-planar~$\lambda$-slope drawings, or \emph{$\lambda$-drawings} for short and refer to the problem of finding these drawings as \ldr.
Since the possible edge slopes in a~$\lambda$-drawing are integers all vertices lie on the integer grid.

\paragraph{Related Work.}
The number of slopes used for the edges in a graph drawing can be seen as an indication of the simplicity of the drawing.
For instance, the measure \emph{edge orthogonality}, which specifies how close a drawing is to an \emph{orthogonal drawing}, where edges are polylines consisting of horizontal and vertical segments only, has been proposed as a measure of aesthetic quality of a graph drawing~\cite{Purchase02}. In a similar spirit, Kindermann et al.~studied the effect reducing the segment complexity on the aesthetics preference of graph drawings and observed that in some cases people prefer drawings using lower segment complexity~\cite{JGAA-474}.   
More generally, the use of few slopes for a graph drawing may contribute to the formation of ``Pr\"agnanz'' (``good figure'' in German) of the visualization, that accordingly to the Gestalt law of Pr\"agnanz, or law of simplicity, contributes to the perceived  quality of the visualizations.
This is design principle often guides the visualization of metro maps.
See~\cite{Noellenburg14} for a survey of the existing approaches, most of which generate octilinear layouts of metro maps, and~\cite{Nickel19} for a recent model for drawing more general~$k$-linear metro maps.

Level-planar drawing with few slopes have not been considered in the literature but drawings of undirected graphs with few slopes have been extensively studied.
The (\emph{planar}) \emph{slope number} of a (planar) graph~$G$ is the smallest number~$s$ so that~$G$ has a (planar) straight-line drawing with edges of at most~$s$ distinct slopes.
Special attention has been given to proving bounds on the (planar) slope number of undirected graph classes~\cite{BaratMW06,JGAA-376,DESW07,DSW07,GiacomoLM18,KPPT08,KPP13,KnauerMW14,LenhartLMN13,PachP06}.
%For instance, Dujmovi\'c et al.~ \cite{DESW07} showed that cubic planar graph have planar slope number three with the exception of the edges on the outer face which are represented with bends. More generally, Keszegh  et al. have shown that the planar slope number of planar graphs is the function of the maximum degree~\cite{KPP13}.  On the other hand, Bar\'at et al.~\cite{BaratMW06}, and independently Pach and P\'alv\"olgyi~\cite{PachP06} proved that there exist non-planar graphs with maximum degree~$d$ with the unbounded slope number. For non-planar graphs with maximum degree at most three five slopes are sufficient~\cite{KPPT08}, which goes down to four for connected graphs~\cite{MukkamalaS09}.  Slope number has also been studied for polyline drawings, we refer the reader to~\cite{KindermannMSS18} for a recent overview of the results.
%From the algorithmic perspective, deciding whether an undirected graph has a drawing with only two orthogonal slopes is NP-hard~\cite{FHH+93}, and the same holds for planar setting~\cite{GT01}. %However, for plane graphs of  maximum degree three
%the decision can be done in linear time~\cite{RahmanNN02}. 
%, for series-parallel graph G of the maximum degree three~\cite{RahmanEN05} and 
%for subdividions of planar triconnected cubic graphs~\cite{RahmanEN05} 
Determining the planar slope number is hard in the existential theory of reals~\cite{Hoffmann17}. 

%If the edges are allowed to have bends and to only consist of vertical and horizontal segments we talk about  \emph{orthogonal drawings}.  The problem of bend minimization problem in planar orthogonal drawings is NP-hard~\cite{GargT01}. However, back in 1987, Tamassia presented an polynomial time algorithm that
%performs bend minimization for plane graphs~\cite{Tam87}. 
%The running time was subsequently improved  to~$O(n^{3/2} \log n)$~\cite{CK12}. This algorithm is based on constructing a flow network with lower bounds and capacities on the edges and showing that a minimum cost  flow  in this network corresponds to a drawing with the minimum number of bends. 

Several graph visualization problems have been considered in the partial and simultaneous settings.
%These formulations are motivated by temporal graphs, user interaction and their intrinsic theoretic complexity.
In the \emph{partial drawing extension} problem, one is presented with a graph and an immutable drawing of some subgraph thereof.
The task is to determine whether the given drawing of the subgraph can be completed to a drawing of the entire graph. The problem has been studied for the planar setting~\cite{MNR16,Patrignani06} and also the level-planar setting~\cite{BR17}.
In the \emph{simultaneous drawing} problem, one is presented with two graphs that may share some subgraph.
The task is to draw both graphs so that the restrictions of both drawings to the shared subgraph are identical. We refer the reader to~\cite{BlasiusKR13} for an older literature overview.
The problem has been considered  for orthogonal drawings~\cite{ACC+16} and level-planar drawings~\cite{ADLDB+16}.  Up to our knowledge, neither partial nor simultaneous drawings have been considered in the restricted slope setting.
%Frati et al.\ studied the setting where the embedding is fixed~\cite{FKK09}.

\paragraph{Contribution.}
We introduce and study the \ldr problem.
To solve this problem for proper level graphs, we introduce two models.
In Section~\ref{sec:flow-model} we describe the first model, which uses a classic integer-circulation-based approach.
This model allows us to solve the \ldr in~$O(n \log^3 n)$ time and obtain a~$\lambda$-drawing within the same running time if one exists.
In Section~\ref{sec:distance-model}, we describe the second distance-based model.
It uses the duality between flows in the primal graph and distances in the dual graph and allows us to solve the \ldr in~$O(n \log^2 n / \log\log n)$ time.
We also address the partial and simultaneous settings.
The classic integer-circulation-based approach can be used to extend connected partial~$\lambda$-drawings in~$O(n \log^3 n)$ time.
In Section~\ref{sec:partial-and-simultaneous-drawings}, we build on the distance-based model to extend not-necessarily-connected partial~$\lambda$-drawings in~$O(n^{4/3} \log n)$ time, and to obtain simultaneous~$\lambda$-drawings in~$O(\lambda n^{10/3} \log n)$  time if they exist.
We finish with some concluding remarks in Section~\ref{sec:conclusion} and refer to Appendix~\ref{sec:general-case} for a proof that 2-\textsc{Drawability} is \textsf{NP}-hard even for biconnected graphs where all edges have length one or two.

\section{Preliminaries}
\label{sec:preliminaries}
Let~$\Gamma$ be a level-planar drawing of an embedded level-planar graph~$\cal G$.
The \emph{width} of~$\Gamma$ is defined as~$\max_{v \in V} \Gamma(v) - \min_{v \in V} \Gamma(v)$.
An integer~$\bar x$ is a \emph{gap} in~$\Gamma$ if it is~$\Gamma(v) \neq \bar x$ for all~$v \in V$,~$\Gamma(v_1) < \bar x$ and~$\Gamma(v_2) > \bar x$ for some~$v_1, v_2 \in V$, and~$\Gamma(u) < \bar x < \Gamma(v)$ for no edge~$(u, v) \in E$.
A drawing~$\Gamma$ is \emph{compact} if it has no gap.
Note that a~$\lambda$-drawing of a connected level-planar graph is inherently compact.
While in a~$\lambda$-drawing of a non-connected level-planar graph every gap can be eliminated by a shift. 
The fact that we only need to consider compact~$\lambda$-drawings helps us to limit their width. Thus, any compact~$\lambda$-drawing has a width of at most~$(\lambda - 1)(n - 1)$.

Let~$u$ and~$v$ be two vertices on the same level~$i$.
With~$[u, v]_{\mathcal G}$ (or~$[u, v]$ when~$\mathcal G$ is clear from the context) we denote the set of vertices that contains~$u, v$ and all vertices in between~$u$ and~$v$ on level~$i$ in~$\mathcal G$.
We say that two vertices~$u$ and~$v$ are \emph{consecutive in~$\mathcal G$} when~$[u, v]=\{u,v\}$.
Two edges~$e = (u, w), e' = (v, x)$ are \emph{consecutive in~$\mathcal G$} when the only edges with one endpoint in~$[u, v]_{\mathcal G}$ and the other endpoint in~$[w, x]_{\mathcal G}$ are~$e$ and~$e'$.

A \emph{flow network}~$F = (N,A)$ consists of a set of nodes~$N$ connected by a set of directed arcs~$A$.
Each arc has a \emph{demand} specified by a function~$d : A \to \mathbb N_0$ and a \emph{capacity} specified by a function~$c: A \to \mathbb N \cup \{\infty\}$ where~$\infty$ encodes unlimited capacity.
A \emph{circulation} in~$F$ is a function~$\varphi: A \to \mathbb N_0$ that assigns an integral flow to each arc of~$F$ and satisifies the two following conditions.
First, the circulation has to respect the demands and capacities of the arcs, i.e., for each arc~$a \in A$ it is~$d(a) \le \varphi(a) \le c(a)$.
Second, the circulation has to respect flow conservation, i.e., for each node~$v \in N$ it is~$\sum_{(u, v) \in A} \varphi(u, v) = \sum_{(v, u) \in A} \varphi(v, u)$.
Depending on the flow network no circulation may exist.

\section{Flow Model}
\label{sec:flow-model}

\begin{figure}[t]
    \centering
    \includegraphics[width=\linewidth]{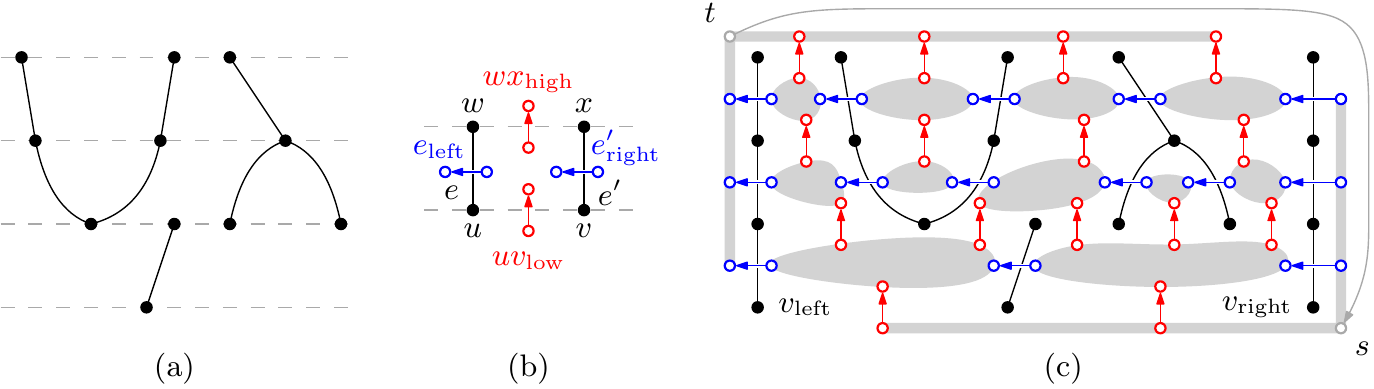}
    \caption{%
        (a) An embedded level graph~$\mathcal G$. (b) The definition of the arcs of the flow network. (c) The graph~$\mathcal G$ together with the paths~$p_{\Left}$ and~$p_{\Right}$ in black. The resulting flow network~$F_{\mathcal G}^\lambda$~(c) consists of the blue slope arcs and the red space arcs, its nodes are formed by merging the nodes in the gray areas.
        The red space arcs have a demand of~$1$ and a capacity of~$(\lambda - 1)(n - 1)$ and the blue slope arcs have a demand of zero and a capacity of~$\lambda - 1$.
    }
    \label{fig:flow-network}
\end{figure}

In this section, we model the \ldr  as a problem of finding a circulation in a flow network.
Let~$\mathcal G$ be an embedded proper~$k$-level graph together with a  level-planar drawing~$\Gamma$.
As a first step, we add two directed paths~$p_{\Left}$ and~$p_{\Right}$ that consist of one vertex on each level from 1 to~$k$ to~$\mathcal G$.
Insert~$p_{\Left}$ and~$p_{\Right}$ into~$\Gamma$ to the left and right of all other vertices as the \emph{left} and \emph{right boundary}, respectively. 
%The purpose of these paths is to be a common point of reference when dealing with the relative positions of vertices extracted from the flow network and to make the definition of the source and sink of the integer flow network easier.
See Fig.~\ref{fig:flow-network}~(a) and (c).
From now on, we assume that~$\mathcal G$ and~$\Gamma$ contain the left and right boundary.

The flow network~$F_{\mathcal G}^\lambda$ consists of nodes and arcs and is similar to a directed dual of~$\mathcal G$ with the difference that it takes the levels of~$\mathcal G$ into account.
In particular, for every edge~$e$ of~$\mathcal G$,~$F_{\mathcal G}^\lambda$  contains two nodes~$e_{\Left}$ and~$e_{\Right}$, in the left and the right faces incident to~$e$, and a dual \emph{slope arc}~$e^\star = (e_{\Right}, e_{\Left})$ with demand~$0$ and capacity~$\lambda - 1$; see the blue arcs in Fig.~\ref{fig:flow-network}~(b) and~(c).
The flow across~$e^\star$ determines the slope of~$e$.
Additionally, for every pair of consecutive vertices~$u, v$ we add two nodes~$[u, v]_{\low}$ and~$[u, v]_{\high}$ to~$F_{\mathcal G}^\lambda$ and connect them by a \emph{space arc}~$[u, v]^\star$; see the red arcs in Fig.~\ref{fig:flow-network}~(b) and~(c).
The flow across~$[u, v]^\star$ determines the space between~$u$ and~$v$.
The space between~$u$ and~$v$ needs to be at least one to prevent~$u$ and~$v$ from colliding and can be at most~$(\lambda - 1)(n - 1)$ due to the restriction to compact drawings.
So, assign to~$[u, v]^\star$ a demand of one and a capacity of~$(\lambda - 1)(n - 1)$.
To obtain the final flow network we merge certain nodes.
Let~$e = (u, w)$ and~$e' = (v, x)$ be consecutive edges.
Merge the nodes~$e_{\Right}, e'_{\Left}$,  the nodes~$\{ \{u',v'\}_{\high} : \forall u', v' \textrm{~consecutive in~}[u, v]\}$  and the nodes
$\{ \{w', x'\}_{\low} : \forall w', x' \textrm{~consecutive in~}[w, x]\}$ into a single node.
Next, merge all remaining source and sink nodes into one source node~$s$ and one sink node~$t$, respectively.
See Fig.~\ref{fig:flow-network}~(c), where the gray areas touch nodes that are merged into a single node.
Finally, insert an arc from~$t$ to~$s$ with unlimited capacity.

The network~$F_{\mathcal G}^\lambda$ is designed in such a way that the circulations in~$F_{\mathcal G}^\lambda$ correspond bijectively to the~$\lambda$-drawings of~$\mathcal G$.
Let~$\Gamma$ be a drawing of~$\mathcal G$ and let~$x$ be the function that assigns to each vertex of~$\mathcal G$ its~$x$-coordinate in~$\Gamma$.
We define a dual circulation~$x^\star$ as follows.
Recall that every arc~$a$ of~$F_{\mathcal G}^\lambda$ is either dual to an edge of~$\mathcal G$ or to two consecutive vertices in~$\mathcal G$.
Hence, the left and right incident faces~$f_{\Left}$ and~$f_{\Right}$ of~$a$ in~$F_{\mathcal G}^\lambda$ contain vertices of~$\mathcal G$.
Define the circulation~$x^\star$ by setting~$x^\star(a) := x(f_{\Right}) - x(f_{\Left})$.
We remark the following, although we defer the proof to the next section.

\begin{lemma}
    Let~$\mathcal G$ be an embedded proper level-planar graph together with a~$\lambda$-drawing~$\Gamma$.
    The dual~$x^\star$ of the function~$x$ that assigns to each vertex of~$\mathcal G$ its~$x$-coordinate in~$\Gamma$ is a circulation in~$F_{\mathcal G}^\lambda$.
    \label{lem:drawing-to-circulation}
\end{lemma}

In the reverse direction, given a circulation~$\varphi$ in~$F_{\mathcal G}^\lambda$ we define a dual function~$\varphi^\star$ that, when interpeted as assigning an~$x$-coordinate to the vertices of~$\mathcal G$, defines a~$\lambda$-drawing of~$G$.
Refer to the level-1-vertex~$p_{\Right}$ as~$v_{\Right}$.
Start by setting~$\varphi^\star(v_{\Right}) = 0$, i.e., the~$x$-coordinate of~$v_{\Right}$ is~0.
Process the remaining vertices of the right boundary in ascending order with respect to their levels.
Let~$(u, v)$ be an edge of the right boundary so that~$u$ has already been processed and~$v$ has not been processed yet.
Then set~$\varphi^\star(v) = \varphi^\star(u) + \varphi((u, v)^\star)$, where~$(u, v)^\star$ is the slope arc dual to~$(u, v)$.
Let~$w, x$ be a pair of consecutive vertices so that~$x$ has already been processed and~$w$ has not yet been processed yet.
Then set~$\varphi^\star(w) = \varphi^\star(x) + \varphi([w, x]^\star)$, where~$[w, x]^\star$ is a space arc.
It turns out that~$\varphi^\star$ defines a~$\lambda$-drawing of~$\mathcal G$.

\begin{lemma}
    Let~$\mathcal G$ be an embedded proper level-planar graph,  let~$\lambda \in {\mathbb N}$ and let~$\varphi$ be a circulation in~$F_{\mathcal G}^\lambda$.
    Then the dual~$\varphi^\star$, when interpeted as assigning an~$x$-coordinate to the vertices of~$\mathcal G$, defines a~$\lambda$-drawing of~$G$.
    \label{lem:circulation-to-drawing}
\end{lemma}

While  both  Lemma~\ref{lem:drawing-to-circulation} and Lemma~\ref{lem:circulation-to-drawing} can be proven directly, we defer their proofs to Section~\ref{sec:distance-model} where we introduce the distance model and prove Lemma~\ref{lem:drawing-to-distance-labeling-to-circulation} and Lemma~\ref{lem:circulation-to-distance-labeling-to-drawing}, the stronger versions of Lemma~\ref{lem:drawing-to-circulation} and Lemma~\ref{lem:circulation-to-drawing}, respectively.
Combining  Lemma~\ref{lem:drawing-to-circulation} and  Lemma~\ref{lem:circulation-to-drawing} we obtain the following.

\begin{theorem}
    Let~$\mathcal G$ be an embedded proper level-planar graph and let~$\lambda \in {\mathbb N}$.
    The circulations in~$F_{\mathcal G}^\lambda$ correspond bijectively to the~$\lambda$-drawings of~$\mathcal G$.
    \label{theorem:flow}
\end{theorem}

Theorem~\ref{theorem:flow} implies that a~$\lambda$-drawing can be found by applying existing flow algorithms to~$F_{\mathcal G}^\lambda$. For that, we transform our flow network with arc demands to the standard maximum flow setting without demands by introducing new sources and sinks.
We can then use the~$O(n \log^3 n)$-time multiple-source multiple-sink maximum flow algorithm due to Borradaile et al.~\cite{BKM+11} to find a circulation in~$F_{\mathcal G}^\lambda$ or to determine that no circulation exists.

\begin{corollary}
    Let~$\mathcal G$ be an embedded proper level-planar graph and let~$\lambda \in {\mathbb N}$.
    It can be tested in~$O(n \log^3 n)$ time whether a~$\lambda$-drawing of~$\mathcal G$ exists, and if so, such a drawing can be found within the same running time.
\end{corollary}

\subsection{Connected Partial Drawings}
\label{sec:connected partial}
Recall that a partial~$\lambda$-drawing is a tuple~$(\mathcal G, \mathcal H, \Pi)$, where~$\mathcal G$ is an embedded level-planar graph,~$\mathcal H$ is an embedded subgraph of~$\mathcal G$ and~$\Pi$ is a~$\lambda$-drawing of~$\mathcal H$.   We say that~$(\mathcal G, \mathcal H, \Pi)$ is \emph{$\lambda$-extendable} if~$\mathcal G$ 
admits a~$\lambda$-drawing~$\Gamma$ whose restriction to~$\mathcal H$ is~$\Pi$. Here~$\Gamma$ is referred to as a \emph{$\lambda$-extension} of~$(\mathcal G, \mathcal H, \Pi)$.

In this section we show that in case~$\mathcal H$ is connected, we can use the flow model to decide whether~$(\mathcal G, \mathcal H, \Pi)$ is~$\lambda$-extendable. Observe that when~$\mathcal H$ is connected~$\Pi$ is completely defined by the slopes of the edges in~$\mathcal H$ up to horizontal translation.
Let~$F_{\mathcal G}^\lambda$ be the flow network corresponding to~$\mathcal G$.
In order to fix the slopes of an edge~$e$ of~$\mathcal H$ to a value~$\ell$, we fix the flow across the dual slope arc~$e^\star$  in~$\mathcal H$ to~$\ell$. Checking whether a circulation in the resulting flow network exists can be reduced to a multiple-source multiple-sink maximum flow problem, which once again can be solved by the algorithm due to Borradaile et al.~\cite{BKM+11}.

\begin{corollary}
    Let~$(\mathcal G, \mathcal H, \Pi)$ be a partial~$\lambda$-drawing  where~$\mathcal H$ is connected.
    It can be tested in~$O(n \log^3 n)$ time whether~$(\mathcal G, \mathcal H, \Pi)$ is~$\lambda$-extendable, and if so, a corresponding~$\lambda$-extension  can be constructed within the same running time.
\end{corollary}

\section{Dual Distance Model}
\label{sec:distance-model}

A minimum cut (and, equivalently, the value of the maximum flow) of an~$st$-planar graph~$G$ can be determined by computing a shortest~$(s^\star, t^\star)$-path in a dual of~$G$~\cite{Hu69,IS79}.
Hassin showed that to construct a flow, it is sufficient to compute the distances from~$s^\star$ to all other vertices in the dual graph~\cite{Has81}.
To the best of our knowledge, this duality has been exploited only for flow networks with arc capacities, but not with arc demands.
In this section, we extend this duality to arcs with demands.
The resulting dual distance model improves the running time for the~$\lambda$-\textsc{Drawability}, %testing of existence of~$\lambda$-drawings, 
lets us test the existence of~$\lambda$-extensions of partial~$\lambda$-drawings for non-connected subgraphs, and allows us to develop an efficient algorithm for testing the existence of %yet-to-be-defined 
simultaneous~$\lambda$-drawings.

\begin{figure}[t]
    \centering
    \includegraphics[width=\linewidth]{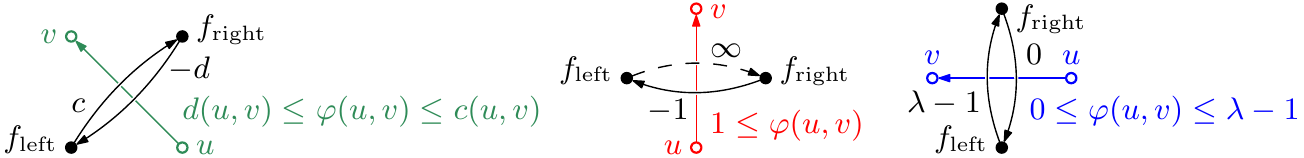}
    \caption{%
        Definition of the dual edges for a flow network arc~$a = (u, v)$ with demand~$d(a)$ and capacity~$c(a)$.
        Let~$f_{\Left}$ and~$f_{\Right}$ denote the vertices corresponding to the faces to the left and right of~$a$ in~$F_{\mathcal G}^\lambda$.
        Then add the edge~$(f_{\Left}, f_{\Right})$ with length~$c(a)$ and the reverse edge~$(f_{\Right}, f_{\Left})$ with length~$-d(a)$.
        Edges with infinite length are not created because they do not add constraints.
    }
    \label{fig:flow-arcs-distance-edges}
\end{figure}

We define~$D_{\mathcal G}^\lambda$ to be the directed dual of~$F_{\mathcal G}^\lambda$ as follows.
Let~$a = (u, v)$ be an arc of~$F_{\mathcal G}^\lambda$ with demand~$d(a)$ and capacity~$c(a)$.
Further, let~$f_{\Left}$ and~$f_{\Right}$ denote the left and the right faces of~$a$ in~$F_{\mathcal G}^\lambda$, respectively.
The dual~$D_{\mathcal G}^\lambda$ contains~$f_{\Left}$ and~$f_{\Right}$ as vertices connected by one edge~$(f_{\Left}, f_{\Right})$ with length~$c(a)$ and another edge~$(f_{\Right}, f_{\Left})$ with length~$-d(a)$; see Fig.~\ref{fig:flow-arcs-distance-edges}.

Observe that to obtain~$F_{\mathcal G}^\lambda$ from~$\mathcal G$ (with left and right paths~$p_{\Left}$ and~$p_{\Right}$) we added dual arcs to edges of~$\mathcal G$ and dual arcs to the space between two consecutive vertices on one level.
Consider for a moment the graph~$\mathcal G'$ obtained from~$\mathcal G$ by adding edges~$(u,v)$ for all consecutive vertices~$u$~$v$, where~$u$ is to the right of~$v$. Graph~$G'$ and ~$D_{\mathcal G}^\lambda$ are identical and therefore ~$D_{\mathcal G}^\lambda$ has the vertex set~$V$ of~$\mathcal G$ and contains a subset of its edges.  
Recall that the dual slope arcs in~$F_{\mathcal G}^\lambda$  have demand~$0$ and capacity~$\lambda-1$, therefore  the edges of ~$D_{\mathcal G}^\lambda$ that connect vertices on different layers have non-negative length. While the edges of~$D_{\mathcal G}^\lambda$  between consecutive vertices on the same level  have  length~$-1$. 

\begin{figure}[t]
    \centering
    \includegraphics{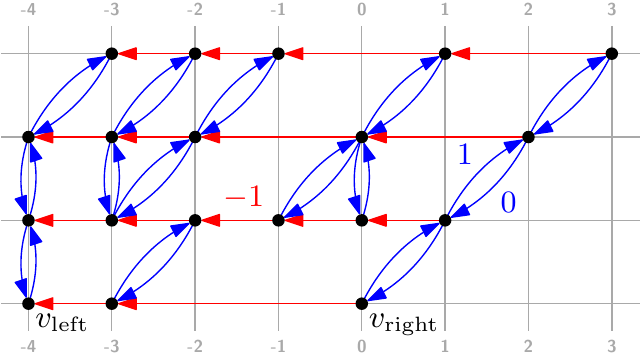}
    \caption{%
        The distance network~$D_{\mathcal G}^2$ obtained from the flow network~$F_{\mathcal G}^2$ shown in Fig.~\ref{fig:flow-network}~(c).
        The~$x$-coordinate of every vertex is its distance from~$v_{\Right}$ in~$D_{\mathcal G}^2$.
        All red arcs have length~$-1$, all blue arcs pointing up have length~$1$ and all blue arcs pointing down have length~$0$.
        For every red arc there exists an arc in the reverse direction with length~$\infty$.
        We omit these arcs because they do not impose any constraints on the shortest distance labeling.
    }
    \label{fig:distance-network}
\end{figure}

A \emph{distance labeling} is a function~$x: V \to \mathbb Z$ that for every edge~$(u, v)$ of~$D_{\mathcal G}^\lambda$ with length~$l$ satisfies~$x(v) \le x(u) + l$.
We also say that~$(u, v)$ \emph{imposes the distance constraint}~$x(v) \le x(u) + l$.
A distance labeling for~$D_{\mathcal G}^\lambda$ is the~$x$-coordinate assignment for a~$\lambda$-drawing:
For an edge~$(u,v)$ of~$D_{\mathcal G}^\lambda$ where~$u,v$ are consecutive vertices in~$\mathcal G$, the distance labeling guarantees~$x(v) \le x(u)-1$, i.e., the consecutive vertices are in the correct order and do not overlap.
If an edge~$(u,v)$ between layers has length~$\lambda - 1$, then the distance labeling ensures~$x(v) \le x(u) + \lambda - 1$, i.e.,~$(u,v)$ has a slope in~$\{0,\dots, \lambda-1\}$.
Computing the shortest distances from~$v_{\Right}$ in~$D_{\mathcal G}^\lambda$ to every vertex (if they are well-defined) gives a distance labeling that we refer to as the \emph{shortest distance labeling}.
A distance labeling of~$D_{\mathcal G}^\lambda$ does not necessarily exist.
This is the case when~$D_{\mathcal G}^\lambda$ contains a negative cycle, e.g., when the in- or out-degree of a vertex in~$\mathcal G$ is strictly larger than~$\lambda$.
For a distance labeling~$x$ of~$D_{\mathcal G}^\lambda$ we define a dual circulation~$x^\star$ by setting~$x^\star(a) := x(f_{\Right}) - x(f_{\Left})$ for each arc~$a$ of~$F_{\mathcal G}^\lambda$ with left and right incident faces~$f_{\Left}$ and~$f_{\Right}$.

\begin{lemma}
    Let~$\mathcal G$ be an embedded level-planar graph and~$\Gamma$ be a~$\lambda$-drawing of~$\mathcal G$.
    The function~$x$ that assigns to each vertex of~$\mathcal G$ its~$x$-coordinate in~$\Gamma$ is a distance labeling of~$D_{\mathcal G}^\lambda$ and its dual~$x^\star$ is a circulation in~$F_{\mathcal G}^\lambda$.
    \label{lem:drawing-to-distance-labeling-to-circulation}
\end{lemma}
\begin{proof}
    Since~$\Gamma$ preserves the embedding of~$\mathcal G$, for each consecutive vertices~$v,~u$, with~$v$ preceeding~$u$ in ~$\mathcal G$  it holds that~$\Gamma(v)<\Gamma(u)$. Since~$\Gamma$ is a grid drawing ~$\Gamma(v) \le \Gamma(u) - 1$, which implies~$x(v)\le x(u) + \ell$, where~$\ell=-1$ is the length of~$(u,v)$. 
    Since  ~$\Gamma$ is a~$\lambda$-drawing, i.e. every edge~$(u,v)$ between the two levels has a slope in~$\{0,\dots \lambda-1\}$, it holds that~$\Gamma(u) \leq \Gamma(v) \le \Gamma(u) + \lambda-1$, which implies~$x(u) \le x(v)+0$, for the edge~$(v,u)$ of~$D_{\mathcal G}^\lambda$ with length zero and~$x(v)<x(u)+\lambda -1$ for the edge~$(u,v)$ of~$D_{\mathcal G}^\lambda$ with length~$\lambda -1$.  
    Hence,~$x$ is a distance labeling of~$D_{\mathcal G}^\lambda$.

    We now show that~$x^\star$ is a circulation in~$F_{\mathcal G}^\lambda$.
    Let~$f_1, f_2, \dots, f_t, f_{t + 1} = f_1$ be the faces incident to some node~$v$ of~$F_{\mathcal G}^\lambda$ in counter-clockwise order.
    Let~$a$ be the arc incident to~$v$ and dual to the edge between~$f_i$ and~$ f_{i + 1}$ with~$1 \le i \le t$.
    If~$a$ is an incoming arc, it adds a flow of~$x(f_{i + 1}) - x(f_{i})$ to~$v$.
    If~$a$ is an outgoing arc, it removes a flow of~$x(f_i) - x(f_{i + 1})$ from~$v$, or, equivalently, it adds a flow of~$x(f_{i + 1}) - x(f_{i})$ to~$v$.
    Therefore, the flow through~$v$ is~$\sum_i \left(x(f_{i + 1}) - x(f_{i})\right)$.
    This sum cancels to zero, i.e., the flow is preserved at~$v$.
    Recall that the edge~$(f_{\Left}, f_{\Right})$ with length~$c(a)$ in~$D_{\mathcal G}^\lambda$ ensures~$x(f_{\Right}) \le x(f_{\Left}) + c(a)$, which gives~$x^\star(a) \le c(a)$.
    So, no capacities are exceeded.
    Analogously, the edge~$(f_{\Right}, f_{\Left})$ with length~$-d(a)$ in~$D_{\mathcal G}^\lambda$ ensures~$x(f_{\Left}) \le x(f_{\Right}) - d(a)$, which gives~$x^\star(a) \ge d(a)$.
    Hence, all demands are fulfilled and~$x^\star$ is indeed a circulation in~$F_{\mathcal G}^\lambda$.
    \qed
\end{proof}

Recall from Section~\ref{sec:flow-model} that for a circulation~$\varphi$ in~$F_{\mathcal G}^\lambda$ we define a dual drawing~$\varphi^\star$ by setting the~$x$-coordinates of the vertices of~$\mathcal G$ as follows.
For the lowest vertex of the right boundary set~$\varphi^\star(v_{\Right}) = 0$.
Process the remaining vertices of the right boundary in ascending order with respect to their levels.
Let~$(u, v)$ be an edge of the right boundary so that~$u$ has already been processed and~$v$ has not been processed yet.
Then set~$\varphi^\star(v) = \varphi^\star(u) + \varphi((u, v)^\star)$, where~$(u, v)^\star$ is the slope arc dual to~$(u, v)$.
Let~$w, x$ be a pair of consecutive vertices so that~$x$ has already been processed and~$w$ has not yet been processed yet.
Then set~$\varphi^\star(w) = \varphi^\star(x) + \varphi([w, x]^\star)$, where~$[w, x]^\star$ is a space arc.
It turns out that~$\varphi^\star$ is a distance labeling of~$D_{\mathcal G}^\lambda$ and a~$\lambda$-drawing of~$\mathcal G$.

\begin{lemma}
    Let~$\mathcal G$ be an embedded level-planar graph, let~$\lambda \in \mathbb N$, and let~$\varphi$ be a circulation in~$F_{\mathcal G}^\lambda$.
    The dual~$\varphi^\star$ is a distance labeling of~$D_{\mathcal G}^\lambda$ and the drawing induced by interpreting the distance label of a vertex as its~$x$-coordinate is a~$\lambda$-drawing of~$\mathcal G$.
    \label{lem:circulation-to-distance-labeling-to-drawing}
\end{lemma}
\begin{proof}
    We show that~$\varphi^\star$ is a distance labeling in~$D_{\mathcal G}^\lambda$.
    The algorithm described above assings a value to every vertex of~$D_{\mathcal G}^\lambda$.
    We now show that~$\varphi^\star$ is indeed a distance labeling by showing that every edge satisfies a distance constraint.

    Observe that the distance constraints imposed by edges dual to the space arcs are satisfied by construction.
    To show that the distance constraints imposed by edges dual to the slope arcs are also satisfied, we prove that for every edge~$(u,v)$, it holds that~$\varphi^\star(v) = \varphi^\star(u) + \varphi((u, v)^\star)$.
    We refer to this as \emph{condition~$\mathcal C$} for short.
    Since~$\varphi((u, v)^\star) \le \lambda -1$ and the length~$\ell$ of~$(u,v)$ is~$\lambda -1$ we obtain~$\varphi^\star(v) = \varphi^\star(u) + \ell$, which implied that~$\phi^\star$ is a distance labeling of~$D_{\mathcal G}^\lambda$.

    The proof is  by induction based on the bottom to top  and  right to left order among the edges of~$D_{\mathcal G}^\lambda$. We say that~$(a,b)$ \emph{precedes}~$(c,d)$ if either~$\ell(a)<\ell(c)$, or~$\ell(a)=\ell(c)$ and~$a$ is to the right of~$c$, or~$\ell(a)=\ell(c)$ and~$b$ is to the right of~$d$ (in case~$a=c$).
    For the base case observe that the edges with both end-vertices on the first level and the edges of~$p_{\Right}$ satisfy condition~$\mathcal C$ by the definition of~$\varphi ^\star$.
    Now let~$(u,v)$ be an edge not addressed in the base case and assume that for every edge~$(u',v')$ preceding edge~$(u,v)$ condition~$\mathcal C$ holds.
    For the inductive step we show that condition~$\mathcal C$ also holds for~$(u,v)$.
    Let~$(u', v')$ denote the edge to the right of~$(u, v)$ so that~$(u, v)$ and~$(u', v')$ are consecutive; see Fig.~\ref{fig:circulation-to-distance-labeling-to-drawing}.
    \begin{figure}[t]
        \centering
        \includegraphics{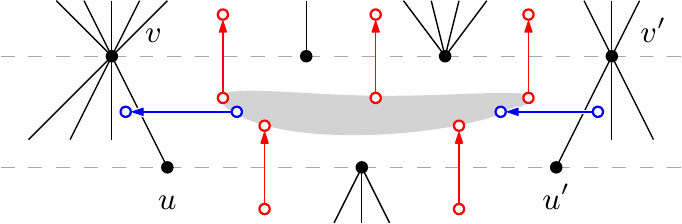}
        \caption{
            Proof of Lemma~\ref{lem:circulation-to-distance-labeling-to-drawing}.
            Sets~$A$ and~$B$ contain the outgoing and incoming red flow network arcs incident to the gray oval, respectively.
        }
        \label{fig:circulation-to-distance-labeling-to-drawing}
    \end{figure}
    Because~$v$ is not the rightmost vertex on its level this edge exists.
    Let~$A$ denote the set of space arcs~$v_1v_2^\star$ in~$F_{\mathcal G}^\lambda$ with~$v_1, v_2 \in [v', v]$.
    Analogously, let~$B$ denote the set of space arcs~$u_1u_2^\star$ in~$F_{\mathcal G}^\lambda$ with~$u_1, u_2 \in [u', u]$.
    It is~$\varphi^\star(v) = \varphi^\star(v') + \sum_{a \in A} \varphi(a)$ by definition of~$\varphi ^\star$.
    Further, by induction hypothesis and since~$(u',v')$ precedes~$(u,v)$ it holds that~$\varphi^\star(v') = \varphi^\star(u') + \varphi((u', v')^\star)$.
    Inserting the latter into the former equation, we obtain~$\varphi^\star(v) = \varphi^\star(u') + \varphi((u', v')^\star) + \sum_{a \in A} \varphi(a)$.
    Again, by definition of~$\varphi ^\star$, it is~$\varphi^\star(u) = \varphi^\star(u') + \sum_{b \in B} \varphi(b)$. By subtracting~$\varphi^\star(u)$ from~$\varphi^\star(v)$ we obtain
     \begin{equation} \varphi^\star(v) = \varphi^\star(u) - \sum_{b \in B} \varphi(b) + \varphi((u', v')^\star) + \sum_{a \in A} \varphi(a) \label{eq:1}\end{equation}
    Flow conservation on the vertex of~$F_{\mathcal G}^\lambda$ to which edges of~$A$ and~$B$ are incident gives ~$\varphi((u, v)^\star) - \sum_{a \in A} \varphi(a) - \varphi((u', v')^\star) + \sum_{b \in B} \varphi(b) = 0$.
    Solving this equation for~$\varphi(u, v)$  and inserting it into~(\ref{eq:1}) yields~$\varphi^\star(v) = \varphi^\star(u) + \varphi((u, v)^\star)$, i.e. the condition~$\mathcal C$ holds for~$(u,v)$. Therefore~$\varphi^\star$ is a distance labeling, which we have shown to define a~$\lambda$-drawing of~$\mathcal G$.
    \qed
\end{proof}

Because~$D_{\mathcal G}^\lambda$ is planar we can use the~$O(n \log^2 n / \log\log n)$-time shortest path algorithm due to Mozes and Wulff-Nilsen~\cite{MWN10} to compute the shortest distance labeling.
This improves our~$O(n \log^3 n)$-time algorithm from Section~\ref{sec:flow-model}.

\begin{theorem}
    Let~$\mathcal G$ be an embedded proper level-planar graph.
    The distance labelings of~$D_{\mathcal G}^k$ correspond bijectively to the~$\lambda$-drawings of~$\mathcal G$.
    If such a drawing exists, it can be found in~$O(n \log^2 n / \log\log n)$ time.
    \label{thm:bijection-distance-labelings-drawings}
\end{theorem}

\section{Partial and Simultaneous Drawings}
\label{sec:partial-and-simultaneous-drawings}

In this section we use the distance model from Section~\ref{sec:distance-model} to construct partial  and simultaneous ~$\lambda$-drawings.
We start with introducing a useful kind of drawing.
Let~$\Gamma$ be a~$\lambda$-drawing of~$\mathcal G$.
We call~$\Gamma$ a \emph{$\lambda$-rightmost} drawing when there exists no~$\lambda$-drawing~$\Gamma'$ with~$\Gamma(v) < \Gamma'(v)$ for some~$v \in V$.
In this definition, we assume~$x(\Gamma(v_{\Right})) = x(\Gamma'(v_{\Right})) = 0$ to exclude trivial horizontal translations.
Hence, a drawing is rightmost when every vertex is at its rightmost position across all level-planar~$\lambda$-slope grid drawings of~$\mathcal G$.
It is not trivial that a~$\lambda$-rightmost drawing exists, but it follows directly from the definition that if such a drawing exists, it is unique.
The following lemma establishes the relationship between~$\lambda$-rightmost drawings and shortest distance labelings of~$D_{\mathcal G}^\lambda$.
\begin{lemma}
    Let~$\mathcal G$ be an embedded proper level-planar graph.
    If~$D_{\mathcal G}^\lambda$ has a shortest distance labeling it describes the~$\lambda$-rightmost drawing of~$\mathcal G$.
    \label{lem:rightmost-drawing}
\end{lemma}
\begin{proof}
    The shortest distance labeling of~$D_{\mathcal G}^\lambda$ is maximal in the sense that for any vertex~$v$ there exists a vertex~$u$ and an edge~$(u, v)$ with length~$l$ so that it is~$x(v) = x(u) + l$.
    Recall that the definition of distance labelings only requires~$x(v) \le x(u) + l$.
    The claim then follows by induction over~$V$ in ascending order with respect to the shortest distance labeling.
    \qed
\end{proof}

\subsection{Partial Drawings}

Let~$(\mathcal G, \mathcal H, \Pi)$ be a partial~$\lambda$-drawing. 
In Section~\ref{sec:connected partial} we have shown that the flow model  can be adapted  to check whether~$(\mathcal G, \mathcal H, \Pi)$ has a~$\lambda$-extension, in case~$\mathcal H$ is connected.  In this section, we show how to adapt the distance model  to extend partial~$\lambda$-drawings, including the case~$\mathcal H$ is disconnected.
Recall that the distance label of a vertex~$v$ is its~$x$-coordinate.
A partial~$\lambda$-drawing fixes the~$x$-coordinates of the vertices of~$\mathcal H$.
The idea is to express this with additional constraints in~$D_{\mathcal G}^\lambda$.
Let~$v_{\refv}$ be a vertex of~$\mathcal H$. In a~$\lambda$-extension of~$(\mathcal G, \mathcal H, \Pi)$, the relative distance along the~$x$-axis between a vertex~$v$ of~$\mathcal H$ and vertex~$v_{\refv}$ should be~$d_v=\Pi(v_{\refv})-\Pi(v)$. 
This can be achieved by adding an edge~$(v, v_{\refv})$ with length~$d_v$ and an edge~$(v_{\refv}, v)$ with length~$-d_v$.
The first edge ensures that it is~$x(v_{\refv}) \le x(v) + d_v$, i.e.,~$x(v) \ge x(v_{\refv}) - d_v$ and the second edge ensures~$x(v) \le x(v_{\refv}) - d$.
Together, this gives~$x(v) = x(v_{\refv}) - d_v$.
Let~$D_{{\mathcal G},\Pi}^\lambda$ be~$D_{\mathcal G}^\lambda$ augmented by the edges~$\{(v, v_{\refv}), (v_{\refv,v}): \forall v \in {\mathcal H}\}$ with lengths as described above.

%See Figure~\ref{fig:distance-network-partial}~(a).
%\begin{figure}
%    \centering
%    \includegraphics[page=1]{figures/distance-network-partial}
%    \hfil
%    \includegraphics[page=2]{figures/distance-network-partial}
%   \caption{
%        Extending partial grid drawings.
%        In~(a), vertex~$u$ is fixed to the~$x$-coordinate~$0$ by the introducing of the two green edges.
%        In~(b), vertex~$w$ should be fixed to the~$x$-coordinate~$-1$.
%        This introduces the negative cycle consisting of bold edges, i.e., this partial drawing cannot be extended.
%    }
%    \label{fig:distance-network-partial}
%\end{figure}

To decide existence of~$\lambda$-extension and in affirmative construct the corresponding drawing we  compute the shortest distance labeling in~$D_{{\mathcal G},\Pi}^\lambda$. Observe that this network can contain negative cycles and therefore no shortest distance labeling.  Unfortunately,~$D_{{\mathcal G},\Pi}^\lambda$ is not planar, and thus we cannot use the embedding-based algorithm of Mozes and Wulff-Nilsen.
However, since all newly introduced edges have~$v_{\refv}$ as one endpoint,~$v_{\refv}$ is an \emph{apex} of~$D_{\mathcal G}^\lambda$, i.e., removing~$v_{\refv}$ from~$D_{{\mathcal G},\Pi}^\lambda$ makes it planar.
Therefore~$D_{{\mathcal G},\Pi}^\lambda$ can be recursively separated by separators of size~$O(\sqrt n)$.
We can therefore use the shortest-path algorithm due to Henzinger et al.\ to compute the shortest distance labeling of~$D_{{\mathcal G},\Pi}^\lambda$ in~$O(n^{4/3} \log n)$ time~\cite{HKRS97}.

\begin{theorem}
    Let~$(\mathcal G, \mathcal H, \Pi)$ be a partial~$\lambda$-drawing.
    In~$O(n^{4/3} \log n)$  time it can be determined whether~$(\mathcal G, \mathcal H, \Pi)$ has a~$\lambda$-extension and in the affirmative the corresponding drawing can be computed within the same running time.
\end{theorem}

\subsection{Simultaneous Drawings}
\label{ssec:simultaneous-drawings}

In the simultaneous~$\lambda$-drawing problem, we are given a tuple~$(\mathcal G_1, \mathcal G_2)$ of two embedded level-planar graphs that share a common subgraph~$\mathcal G_{1 \cap 2} = \mathcal G_1 \cap \mathcal G_2$.
We assume w.l.o.g.\ that~$G_1$ and~$G_2$ share the same right boundary and that the embeddings of~$\mathcal G_1$ and~$\mathcal G_2$ coincide on~$\mathcal G_{1 \cap 2}$.
The task is to determine whether there exist~$\lambda$-drawings~$\Gamma_1, \Gamma_2$ of~$\mathcal G_1, \mathcal G_2$, respectively, so that~$\Gamma_1$ and~$\Gamma_2$ coincide on the shared graph~$\mathcal G_{1 \cap 2}$.
The approach is the following.
Start by computing the rightmost drawings of~$\mathcal G_1$ and~$\mathcal G_2$.
Then, as long as these drawings do not coincide on~$\mathcal G_{1 \cap 2}$ add necessary constraints to~$D_{\mathcal G_1}^\lambda$ and~$D_{\mathcal G_2}^\lambda$.
This process terminates after a polynomial number of iterations, either by finding a simultaneous~$\lambda$-drawing, or by determining that no such drawing exist.

Finding the necessary constraints works as follows.
Suppose that~$\Gamma_1, \Gamma_2$ are the rightmost drawings of~$\mathcal G_1, \mathcal G_2$, respectively.
Because both~$\mathcal G_1$ and~$\mathcal G_2$ have the same right boundary they both contain vertex~$v_{\Right}$.
We define the coordinates in the distance labelings of~$D_{\mathcal G_1}^\lambda$ and~$D_{\mathcal G_2}^\lambda$ in terms of this reference vertex.

Now suppose that for some vertex~$v$ of~$\mathcal G_{1 \cap 2}$ the~$x$-coordinates in~$\Gamma_1$ and~$\Gamma_2$ differ, i.e., it is~$\Gamma_1(v) \neq \Gamma_2(v)$.
Assume~$\Gamma_1(v) < \Gamma_2(v)$ without loss of generality.
Because~$\Gamma_1$ is a rightmost drawing, there exists no drawing of~$\mathcal G_1$ where~$v$ has an~$x$-coordinate greater than~$\Gamma_1(v)$.
In particular, there exist no simultaneous drawings where~$v$ has an~$x$-coordinate greater than~$\Gamma_1(v)$.
Therefore, we must search for a simultaneous drawing where~$\Gamma_2(v) \le \Gamma_1(v)$.
We can enforce this constraint by adding an edge~$(v_{\Right}, v)$ with length~$\Gamma_1(v)$ into~$D_{\mathcal G_2}^\lambda$.
We then attempt to compute the drawing~$\Gamma_2$ of~$\mathcal G_2$ defined by the shortest distance labeling in~$D_{\mathcal G_2}^\lambda$.
This attempt produces one of two possible outcomes.
The first possibility is that there now exists a negative cycle in~$D_{\mathcal G_2}^\lambda$.
This means that there exists no drawing~$\Gamma_2$ of~$G_2$ with~$\Gamma_2(v) \le \Gamma(v)$.
Because~$\Gamma_1$ is a rightmost drawing, this means that no simultaneous drawings of~$\mathcal G_1$ and~$\mathcal G_2$ exist.
The algorithm then terminates and rejects this instance.
The second possiblity is that we obtain a new drawing~$\Gamma_2$.
This drawing is rightmost among all drawings that satisfy the added constraint~$\Gamma_2(v) \le \Gamma_1(v)$.
In this case there are again two possibilities.
Either we have~$\Gamma_1(v) = \Gamma_2(v)$ for each vertex~$v$ in~$\mathcal G_{1 \cap 2}$.
In this case~$\Gamma_1$ and~$\Gamma_2$ are simultaneous drawings and the algorithm terminates.
Otherwise there exists at least one vertex~$w$ in~$\mathcal G_{1 \cap 2}$ with~$\Gamma_1(w) \neq \Gamma_2(w)$.
We then repeat the procedure just described for adding a new constraint.

We repeat this procedure of adding other constraints.
To bound the number of iterations, recall that we only consider compact drawings, i.e., drawings whose width is at most~$(\lambda - 1)(n - 1)$.
In each iteration the~$x$-coordinate of at least one vertex is decreased by at least one.
Therefore, each vertex is responsible for at most~$(\lambda - 1)(n - 1)$ iterations.
The total number of iterations is therefore bounded by~$n (\lambda - 1)(n - 1) \in O(\lambda n^2)$.

Note that due to the added constraints~$D_{\mathcal G_1}^\lambda$ and~$D_{\mathcal G_2}^\lambda$ are generally not planar.
We therefore apply the~$O(n^{4/3} \log n)$-time shortest-path algorithm due to Henzinger et al.~that relies not on planarity but on~$O(\sqrt n)$-sized separators to compute the shortest distance labellings.
This gives the following.

\begin{theorem}
    Let~$\mathcal G_1, \mathcal G_2$ be embedded level-planar graphs that share a common subgraph~$\mathcal G_{1 \cap 2}$.
    In~$O(\lambda n^{10/3} \log n)$ time it can be determined whether~$\mathcal G_1, \mathcal G_2$ admit simultaneous~$\lambda$-drawings and if so, such drawings can be computed within the same running time.
\end{theorem}

\section{Conclusion}
\label{sec:conclusion}

In this paper we studied~$\lambda$-drawings, i.e., level-planar drawings with~$\lambda$ slopes.
We model~$\lambda$-drawings of proper level-planar graphs as integer flow networks.
This lets us find~$\lambda$-drawings and extend connected partial ~$\lambda$-drawings in~$O(n \log^3 n)$ time.
We extend the duality between integer flows in a primal graph and shortest distances in its dual to obtain a more powerful distance model.
This distance model allows us to find~$\lambda$-drawings in~$O(n \log^2 n / \log\log n)$ time, extend not-necessarily-connected partial ~$\lambda$-drawings in~$O(n^{4/3} \log n)$ time and find simultaneous ~$\lambda$-drawings in~$O(\lambda n^{10/3} \log n)$ time.
%Our simultaneous grid drawing algorithm solves a special case of the simultaneous circulation problem which we believe to be of general interest.

In the non proper case, testing the existence of a~$2$-drawing becomes \textsf{NP}-hard, even for  biconnected graphs with maximum edge length two; see Appendix~\ref{sec:general-case}.
%This leaves little room to extend our polynomial-time algorithms for more general classes of level-planar graphs.

%\bibliographystyle{plain}
%\bibliography{references}

\newpage
\appendix

\section{Complexity of the General Case}
\label{sec:general-case}

So far, we have considered proper level graphs, i.e., level graphs where all edges have length one.
In this section, we consider the general case, where edges may have arbitrary lengths.
We say that an edge with length two or more is \emph{long}.
One approach would be to try to adapt the flow model from Section~\ref{sec:flow-model} to this more general case.
By subdividing long edges, any level graph~$G$ can be transformed into a proper level graph~$G'$.
Observe that two edges in~$G'$ created by subdividing the same long edge must have the same slope in order to yield a fixed-slope drawing of~$G$.
In the context of our flow model, this means that the amount of flow across the corresponding slope arcs must be the same.
Our problem then becomes an instance of the \emph{integer equal flow problem}.
In this problem, we are given a flow network along with disjoint sets~$R_1, R_2, \dots, R_t$ of arcs.
The task is to find the maximum flow from~$s$ to~$t$ such that the amount of flow across arcs in the same set~$R_i$ is the same.
This problem was introduced and shown to be {\sf NP}-hard by Sahni~\cite{Sah74}.
The problem remains {\sf NP}-hard in special cases~\cite{EIS76,SGK+02} and the integrality gap of the fractional LP can be arbitrarily large~\cite{MS09}.
%Heuristics have been developed for the case when~$\lvert R_i\rvert = 2$ for~$1 \le i \le t$~\cite{AKS88,LL97}.
%In our setting, this corresponds to the case when all edges have length at most two.

\begin{figure}[b]
    \centering
    \includegraphics{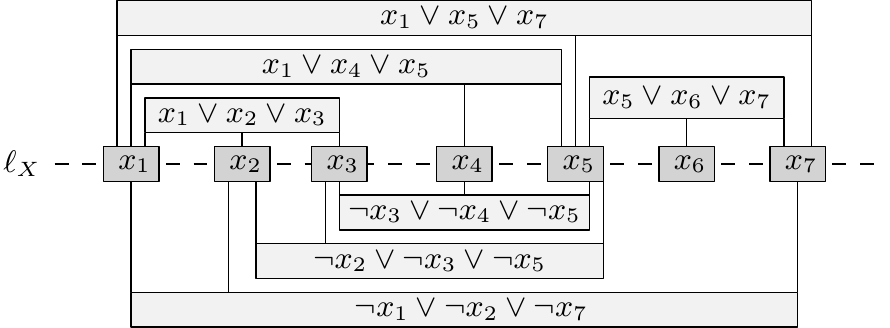}
    \caption{%
        An instance of planar monotone 3-\textsc{Sat}.
    }
    \label{fig:planar-monotone-3sat}
\end{figure}

In this section we show that the level-planar grid drawing problem is {\sf NP}-complete even for two slopes and biconnected graphs where all edges have length one or two.
To this end, we present a reduction from \emph{rectilinear planar monotone 3-\textsc{Sat}}~\cite{dBK10}.
An instance of this problem consists of a set of variables~$X$ and a set of clauses~$C$.
A clause is \emph{positive} (\emph{negative}) when it consists of only positive (negative) literals.
We say that the instance is \emph{monotone} when each clause is either positive or negative.
Further, the instance can be planarly drawn as a graph with vertices~$X \cup C$; see Fig.~\ref{fig:planar-monotone-3sat}.
In this drawing, the variables are aligned along a virtual horizontal line~$\ell_X$.
Positive clauses are drawn as vertices above~$\ell_X$ and connected by an edge to a variable if the clause contains the corresponding postive literal.
Symmetrically, negative clauses are drawn as vertices below~$\ell_X$ and connected by an edge to a variable if the clause contains the corresponding negative literal.

Our reduction works by first replacing every vertex that corresponds to a variable by a variable gadget and every vertex that corresponds to a positive (negative) clause by a positive (negative) clause gadget.
All three gadgets consist of \emph{fixed} and \emph{movable} parts.
The fixed parts only admit one level-planar two-slope grid drawing, whereas the movable parts admit two or more drawings depending on the choice of slope for some edges.
Second, the gadgets are connected by a common fixed frame.
All fixed parts of the gadgets are connected to the common frame in order to provide a common point of reference.
The movable parts of the gadgets then interact in such a way that any level-planar two-slope grid drawing induces a solution to the underlying planar monotone 3-\textsc{Sat} instance.

\begin{figure}[t]
    \centering
    \includegraphics[page=1,width=.49\linewidth]{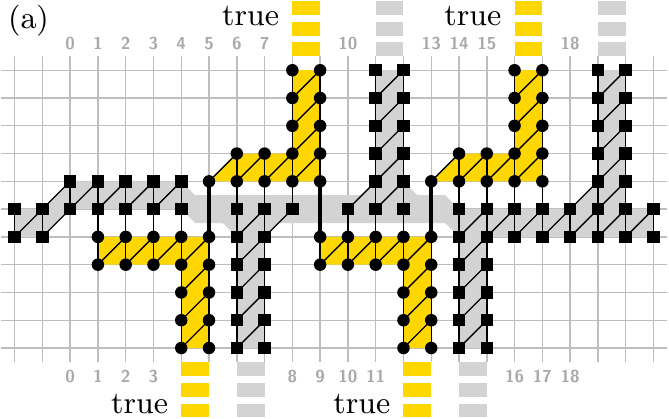}%
    \hfill%
    \includegraphics[page=2,width=.49\linewidth]{figures/variable-gadget}%
    \caption{%
        The variable gadget drawn in the ``true'' configuration~(a) and the ``false'' configuration~(b).
    }
    \label{fig:variable-gadget}
\end{figure}

The variable gadget consists of a number of \emph{connectors} arranged around a fixed horizontal line that connects all variable gadgets along the virtual line of variables~$\ell_X$.
See Fig.~\ref{fig:variable-gadget}, where the fixed structure is shaded in gray.
Vertices drawn as squares are fixed, i.e., they cannot change their position relative to other vertices drawn as squares.
Vertices drawn as circles are movable, i.e., they can change their position relative to vertices drawn as squares.
The line of variables~$\ell_X$ extends from the the square vertices on the left and right boundaries of the drawing.
Every connector consists of two \emph{pins}: the movable \emph{assignment pin} and the fixed \emph{reference pin}.
The variable gadget in Fig.~\ref{fig:variable-gadget} features four connectors: two above the horizontal line and two below the horizontal line.
Assignment pins are shaded in yellow and reference pins are shaded in gray.
The relative position of the assignment pin and the reference pin of one connector encodes the truth assignment of the underlying variable.
Moreover, the reference pin allows the fixed parts of the clause gadgets to be connected to the variable gadgets and thereby to the common frame.
Comparing Fig.~\ref{fig:variable-gadget}~(a) and~(b), observe how the relative position of the two pins of each connector changes depending on the truth assignment of the underlying variable.
The key structure of the variable gadget is that the position of the assignment pins of one variable gadget are coordinated by long edges.
In Fig.~\ref{fig:variable-gadget}, long edges are drawn as thick lines.
Changing the slope of these long edges moves all assignment pins above the horizontal line in one direction and all assignment pins below the horizontal line in the reverse direction.
In this way, all connectors encode the same truth assignment of the underlying variable.
Note that we can introduce as many connectors as needed for any one variable.

\begin{figure}[t]
    \centering
    \includegraphics[page=3]{figures/positive-clause-gadget}\hspace{2\bigskipamount}%
    \includegraphics[page=2]{figures/positive-clause-gadget}

    \bigskip
    \bigskip

    \includegraphics[page=1]{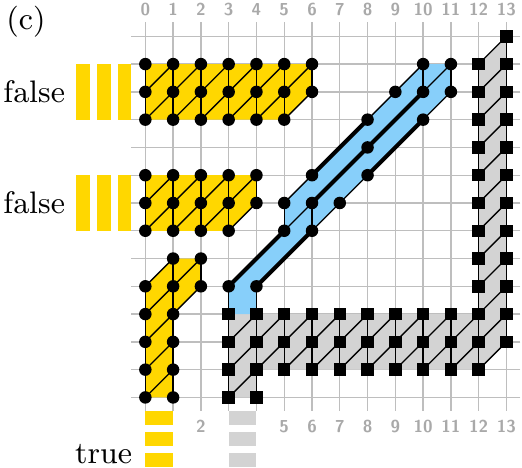}\hspace{2\bigskipamount}%
    \includegraphics[page=4]{figures/positive-clause-gadget}
    \caption{%
        The positive clause gadget.
        Subfigures~(a)--(c) show the drawing when at least one variable is assigned to true.
        Subfigure~(d) shows that no drawing exists when all variables are assigned to false, because this leads to intersections, e.g., at the vertices marked in red.
    }
    \label{fig:positive-clause-gadget}
\end{figure}

The positive clause gadget consists of a fixed boundary, a movable \emph{wiggle} and three \emph{assignment pin endings}.
See Fig.~\ref{fig:positive-clause-gadget}, where the fixed boundary is shaded in gray, the wiggle is highlighted in blue and the assignment pin endings are highlighted in yellow.
The fixed boundary is connected to the reference pin of the variable gadget that is rightmost amongst the connected variable gadgets.
Because the variable gadgets are fixed to the common frame, the boundary of the positive clause gadget is also connected to the common frame.
The assignment pin endings are connected to assigment pins of connectors of the corresponding variable gadgets.
The idea of the positive clause gadget is that the wiggle has to wiggle through the space bounded by the assignment pin endings on the left and the fixed boundary on the right.
Recall that the assignment pins change their horizontal position depending on the truth assignment of their underlying variables.
The positive clause gadget is designed so that the wiggle can always be drawn, except for the case when all variables are assigned to false.
See Fig.~\ref{fig:positive-clause-gadget}~(a)--(c), which shows the three possible situations when exactly one variable is assigned to true.
In any case where at least one variable is assigned to true the wiggle can be drawn in one of the three ways shown.
However, as shown in Fig.~\ref{fig:positive-clause-gadget}~(d), the wiggle cannot be drawn in the case where all variables are assigned to false.
The reason for this is that the the assignment pin endings get so close to the fixed boundary that they leave too little space for the wiggle to be drawn.
This means that some vertices must intersect, for example those shown in red in Fig.~\ref{fig:positive-clause-gadget}~(d).

\begin{figure}[h]
    \centering
    \includegraphics[page=3]{figures/negative-clause-gadget}\hspace{2\bigskipamount}%
    \includegraphics[page=2]{figures/negative-clause-gadget}

    \bigskip
    \bigskip

    \includegraphics[page=1]{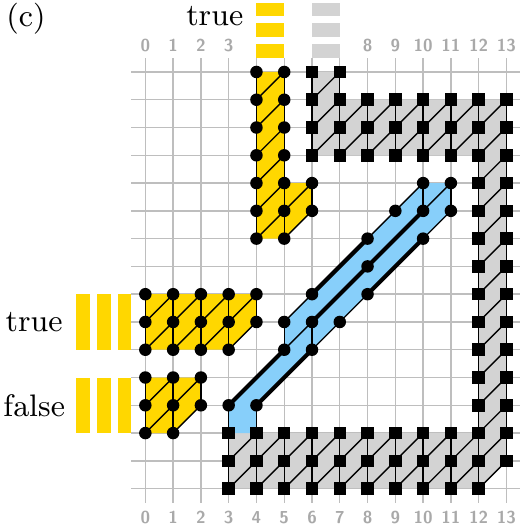}\hspace{2\bigskipamount}%
    \includegraphics[page=4]{figures/negative-clause-gadget}
    \caption{%
        The negative clause gadget.
        Subfigures~(a)--(c) show the drawing when at least one variable is assigned to false.
        Subfigure~(d) shows that no drawing exists when all variables are assigned to true, because this leads to intersections, e.g., at the vertices marked in red.
        Note the symmetry to the positive clause gadget in Fig.~\ref{fig:positive-clause-gadget}.
    }
    \label{fig:negative-clause-gadget}
\end{figure}

The negative clause gadgets works very similarly.
It is drawn below the horizontal line of variables and it forces at least one of the incident variable gadgets to be configured as false.
See Fig.~\ref{fig:negative-clause-gadget}, where subfigures~(a)--(c) show the admissible drawings and subfigure~(d) shows that the case when all incident variables are configured as true cannot occur.
Note that this uses the design of the variable gadget that the assignment and reference pins below the horizontal line are closer when the variable is assigned to true (i.e., the inverse situtation compared to the situation above the horizontal line).

It is evident that any level-planar two-slope grid drawing of the resulting graph induces a solution of the underlying planar monotone 3-\textsc{Sat} problem and vice versa.
Note that the variable gadgets become biconnected when embedded into the common frame and that the clause gadgets are biconnected by design.
Furthermore, all long edges have length two.
We therefore conclude the following.

\begin{theorem}
    The level-planar grid drawing problem is {\sf NP}-complete even for two slopes and biconnected graphs where all edges have length one or two.
\end{theorem}

\end{document}